\documentstyle{article}
\newcommand{\be}{\begin{equation}}
\newcommand{\ee}{\end{equation}}
\newcommand{\bea}{\begin{eqnarray}}
\newcommand{\eea}{\end{eqnarray}}
\newcommand{\bdm}{\begin{displaymath}}
\newcommand{\edm}{\end{displaymath}}
\newcommand{\cod}{d^{\dagger}}

\newcommand{\codga}{d^{\dagger (\gamma)}}
\newcommand{\we}{\wedge}
\newcommand{\ul}{\underline}

\newcommand{\lapgamma}{\Delta^{(\gamma)}}
\newcommand{\lapdelta}{\Delta^{(\delta)}}
\newcommand{\nabdel}{\nabla^{(\delta)}}

\newcommand{\nabiga}{\nabla^{(\gamma)}_i}
\newcommand{\nabjga}{\nabla^{(\gamma)}_j}
\newcommand{\nabzga}{\nabla^{(\gamma)}_z}
\newcommand{\nabrga}{\nabla^{(\gamma)}_{\rho}}

\newcommand{\dx}{\partial_x}
\newcommand{\dy}{\partial_y}

\newcommand{\Ebar}{\overline{E}}
\newcommand{\eform}{{\cal E}}

\newcommand{\Xhat}{\hat{X}}
\newcommand{\Ahat}{\hat{A}}
\newcommand{\Vhat}{\hat{V}}
\newcommand{\What}{\hat{W}}
\newcommand{\hhat}{\hat{h}}
\newcommand{\hhatvac}{\hat{h}^{(vac)}}
\newcommand{\that}{\hat{t}}
\newcommand{\phihat}{\hat{\varphi}}

\newcommand{\Yhat}{\hat{Y}}

\newcommand{\eps}{\epsilon}
\newcommand{\epsbar}{\overline{\epsilon}}

\newcommand{\eebar}{\epsilon \overline{\epsilon}}

\newcommand{\hvac}{h^{(vac)}}
\newcommand{\hphi}{h^{(\phi)}}
\begin{document}
\begin{titlepage}
%
\begin{center}
\vspace*{1.8cm}
\vfill
{\large\bf THE UNIQUENESS THEOREM FOR} \\
{\large\bf ROTATING BLACK HOLE SOLUTIONS OF} \\
{\large\bf SELF-GRAVITATING HARMONIC MAPPINGS}
\vspace{1.2cm}
\vfill
\bf Markus Heusler  \\
\vspace{1.2cm}
\vfill
The Enrico Fermi Institute \\
University of Chicago, 5640 Ellis Avenue \\
Chicago, Illinois 60637 \\
\end{center}
\vfill
\vspace{2.0cm}
\begin{center}
{\bf Abstract}
\end{center}
\begin{quote}
We consider rotating black hole configurations of self-gravitating
maps from spacetime into arbitrary Riemannian manifolds. We first
establish the integrability conditions for the Killing fields generating
the stationary and the axisymmetric isometry (circularity theorem).
Restricting ourselves to mappings with harmonic action, we subsequently
prove that the only stationary and axisymmetric, asymptotically flat
black hole solution with regular event horizon is the Kerr metric.
Together with the uniqueness result for non-rotating configurations and
the strong rigidity theorem, this establishes the uniqueness of the Kerr
family amongst all stationary black hole solutions of selfgravitating
harmonic mappings.
\vspace{3.0cm}
\end{quote}
\vfill
\end{titlepage}
\section{Introduction}

By now, the uniqueness of the $3$-parameter Kerr-Newman
\cite{KN} family amongst all stationary, asymptotically flat
black hole solutions of the coupled Einstein-Maxwell equations
is established quite rigorously. The proof relies heavily on the
fact that the event horizon of a stationary black hole is a
Killing horizon, implying that the null generator Killing field of the
horizon either coincides with the asymptotically timelike
Killing field or the domain of outer communications is
stationary {\em{and}} axisymmetric. By virtue of this
(strong rigidity) theorem \cite{Hawk72b}, \cite{Hawk73},
the stationary black hole configurations are subdivided
into non-rotating and rotating ones.

The uniqueness theorem for the first class of solutions,
that is the uniqueness of the Reissner-Nordstr\"om metric
\cite{RN} amongst all {\em{non-rotating}} electrovac black
hole equilibrium states, was established by the work of M\"uller
zum Hagen, Robinson and Seifert \cite{Mu1}, \cite{Mu2},
Robinson \cite{Robi77} and others, basing on Israel's pioneering
results \cite{Isra67}, \cite{Isra68}. Some open gaps have
been closed more recently, such as the extension of certain
vacuum results to the electrovac case \cite{Simo85}, the
exclusion of multiple black hole solutions \cite{Simo84},
\cite{Bunt87}, \cite{Ruba88}, \cite{Maso92} and the proof
of the staticity theorem for the Einstein-Maxwell system
\cite{Suda92} (see also \cite{Cart87}).

The uniqueness results for {\em{rotating}} configurations,
that is for stationary and axisymmetric black hole
spacetimes, stand on a very solid foundation as well.
Early progress concerning the vacuum case was achieved
by the work of Carter \cite{Cart71}, \cite{Cart73a},
Hawking \cite{Hawk73} and Robinson \cite{Robi74},
\cite{Robi75}. Some years ago, Bunting \cite{Bunt83}
and Mazur \cite{Mazu82} were eventually able to
generalize these results to electromagnetic fields.
(See \cite{Cart85} for a review on the new methods
introduced by these authors.)

The recent discovery of new black hole solutions of the
Einstein-Yang-Mills system (with vanishing Yang-Mills
charges) \cite{Bart88}, \cite{Volk90}, \cite{Volk90a},
\cite{Kunz90} demonstrates that the aforementioned
uniqueness results do {\em{not}} generalize in a
straightforward way in the presence of non-Abelian
gauge fields. The same applies to the related class of
field theories represented by non-linear sigma-models or,
more generally, self-gravitating mappings $\phi$ from
spacetime into Riemannian target manifolds. As a matter
of fact, new black hole solutions with Skyrme hair have
recently been constructed \cite{Droz91} and, in contrast
to the solutions with Yang-Mills hair, turned out to be
(linearly) stable \cite{HDS1}, \cite{HDS2}, \cite{HSZ}.

It is, however, well known that the Skyrme action is not
minimal in the sense that it contains quartic terms in the
differential $d \phi$ of the mapping $\phi$.
(See \cite{Manton} for a generalization of the classical
Skyrme model to arbitrary target spaces.)
Uniqueness theorems for self-gravitating mappings can thus
only exist for a restricted class of mappings. Hence,
one has to answer the following questions: What are the
necessary requirements for (i) the matter action and
(ii) the target manifold $(N,G)$ such that the field equations
admit only the vacuum solution $(g_K,\phi_0)$? (Here $g_K$
and $\phi_0$ denote the Kerr metric and a constant map,
respectively.)

Until recently, only partial answers to the above questions
were known: On the one hand, no-hair theorems were
derived for {\em{static}} mappings into {\em{linear}}
target manifolds, that is,  for the Einstein-Higgs system
(with {\em{convex}} potential) \cite{Sawy77}, \cite{Adle78},
\cite{Zahe86}. On the other hand, it was shown that a static
{\em harmonic} map from a {\em fixed} Schwarzschild
background to a Riemannian manifold has to be constant
\cite{Hu1}, \cite{Hu2}. However, the latter fact does, of
course, not exclude the existence of non-trivial field
configurations in the self-gravitating case. (Consider,
as an example, the corresponding situation for the
Einstein-Yang-Mills system.) To our knowledge, the
strongest results in the {\em{self-gravitating}} case
were obtained for black hole solutions of sigma-models
with harmonic action and non-compact, symmetric
target manifolds $G/H$ \cite{GB} or  (on the basis of
Bochner identities \cite{Bo1}, \cite{Mi1}, \cite{Gi1})
for target manifolds with non-positive sectional curvature.
(In the spherically symmetric case, a generalization of these
results to models with arbitrary Riemannian target manifolds
and arbitrary non-negative potentials was given in
\cite{HScale}, \cite{HJMP} and \cite{HSEC}.)

A complete answer to questions (i) and (ii) was presented
for {\em{non-rotating}} black holes in \cite{HSU}:
It is first shown that any self-gravitating mapping from
a strictly stationary domain of outer communications into
a Riemannian manifold is static. Subsequently, one
demonstrates that the exterior Schwarzschild geometry is
the only maximally extended, static, asymptotically flat
solution of the coupled Einstein-matter equations, provided
that the matter action is {\em harmonic}. This is achieved
by proceeding along the same lines as in the vacuum case
\cite{Bunt87}, making essential use of the positive mass
theorem \cite{Schoen}, \cite{Witten}.

In the present paper we extend the uniqueness result to
the rotating case. We do so by first establishing the
circularity theorem in the second section, which guarantees
the integrability of the $2$-surfaces orthogonal to the Killing
fields generating the stationary and the axisymmetric
isometries. As a consequence, the spacetime metric can be
written in the Papapetrou form \cite{Papa53}. As we shall
argue in the third section, the field equations then split into
two decoupled sets, provided that $\phi$ describes a
self-gravitating {\em{harmonic}} mapping. In the forth
section we show that, like in the vacuum case, both sets
of equations can be derived from a variational principle.
The fifth section is devoted to the first set of equations,
which is exactly the same as in the vacuum case.
Hence, it uniquely determines the Ernst potential
\cite{Ernst68a}, \cite{Ernst68b} (and the associated metric
functions) as the solution of a regular, $2$-dimensional
boundary value problem. The second set of equations
involves the matter fields and an additional metric
function, which is not determined by the Ernst potential.
Using asymptotic flatness and Stoke's theorem for a
suitably chosen $2$-dimensional vector field, we finally
show in the last section that these equations admit
only trivial matter field configurations. This demonstrates
that the only stationary and axisymmetric, asymptotically
flat black hole solutions (with regular event horizon) of
selfgravitating {\em harmonic} mappings consist of the Kerr
metric $g_K$ and constant maps $\phi_0$. Together with
the corresponding result for non-rotating solutions
\cite{HSU} and the strong rigidity theorem, this establishes
the uniqueness of the Kerr family amongst {\em{all}}
stationary black hole configurations with self-gravitating
harmonic mappings.

\section{The Circularity Theorem}

We start this section by briefly recalling the notion of
stationary and axisymmetric, asymptotically flat spacetimes.
Subsequently, we argue that the invariance properties of
the scalar fields with respect to the actions of the two Killing
fields imply that spacetime is Ricci circular. As a consequence,
the circularity theorem \cite{Kund66}, \cite{Cart69} (see also
\cite{Cart87}, \cite{Heus93a}) can be applied in order to
establish the integrability conditions for the Killing fields.
The metric is then written in the Papapetrou form \cite{Papa53} (see also
\cite{Wald84}), representing the starting point for most
investigations on stationary and axisymmetric spacetimes.

In the following we consider self-gravitating mappings
(non-linear sigma-models), described by the action
\be
S \, = \, \int_{M} (\, R \, + \,
{\cal L}(\phi, d \phi, g, G)\, )\, \eta \, ,
\label{1}
\ee
where $\phi$: ($M$,$g$) $\rightarrow$ ($N$,$G$) denotes
a mapping (scalar field) from the spacetime manifold
$M$ with metric $g$ (and volume form $\eta$) to the
{\em{Riemannian}} manifold $N$ with metric $G$.
As far as the {\em circularity} theorem is concerned, the
matter Lagrangian ${\cal L}$ can be an arbitrary expression
in terms of the fields $\phi$, their differentials $d \phi$
and the metrics $g$ and $G$. In order to prove the
{\em uniqueness} theorem we shall, however, have to
restrict ourselves to harmonic Lagrangians,
${\cal L} = \frac{1}{2} \parallel d \phi \parallel ^2$.
The differential $d \phi$ may be considered a section in the
product bundle $\phi^*(TN) \times TM^*$, where $TM^*$
and $\phi^*(TN)$ denote the dual of and the pullback
of the tangential spaces $TM$ and $TN$, respectively
\cite{Misn78}. In local coordinates
of $M$ and $N$ one has the representation
\be
\parallel d \phi \parallel ^2
\, \equiv \, G_{AB} \,
(d \phi^A | d \phi^B) \, = \,
G_{A B}(\phi(x)) \, g^{\mu \nu}(x)
\, \partial_{\mu}
\phi^A \partial_{\nu} \phi^B \, .
\label{2}
\ee
Since ${\cal L}$ does not depend on derivatives of the
spacetime metric, the variation of the matter action
with respect to $g$ does not lead to integrations by parts.
Thus, the energy momentum tensor is given by
\be
T \, = \, 2 \, \frac{\partial {\cal L}}{\partial g^{-1}}
\, - \, {\cal L} \, g \, = \, G_{AB} \, d \phi ^A \, \otimes \,
\frac{\partial {\cal L}}{\partial d \phi _B} \, - \, {\cal L} \, g \, .
\label{3}
\ee

We assume that both spacetime and the matter fields are
{\em{stationary and axisymmetric}}. We recall that an
asymptotically flat spacetime $(M,g)$ is said to be
stationary and axisymmetric if it admits two commuting
Killing fields, $k$ and $m$, where $k$ is asymptotically
timelike and $m$ is spacelike and has closed orbits.
Hence, denoting with $L_K$ the Lie derivative
with respect to an arbitrary vector field $K$, we have
\be
L_k \, g \, = \, L_m \, g \, = \, 0 \, ,
\; \; \; \; \;
L_k \, \phi ^A \, = \, L_m \, \phi ^A \, = \, 0 \, ,
\label{4}
\ee
\be
[ k , m ] \, = \, 0 \, ,
\; \; \; \; \;
X \, \equiv \, (m | m) \, \geq \, 0 \, ,
\; \; \; \; \;
V \, \equiv \, - \, (k | k) \, \geq \, 0 \, ,
\; \; \mbox{asymptotically} \, .
\label{5}
\ee
Note that the requirement that $m$ is everywhere
spacelike or null turns out to be necessary in order to reduce
Einstein's equations to a {\em{regular}} $2$-dimensional
boundary value problem. Besides this, it is implied by
the causality requirement. We also recall that asymptotic
flatness guarantees that the group generated by
$k$ and $m$ is Abelian \cite{Cart70}.

Let us denote with $T(K)$ the energy momentum
$1$-form with respect to an arbitrary Killing field
$K$, $(T(K))_{\mu} \equiv T_{\mu \nu} K^{\nu}$.
If the $\phi^A$ are invariant with respect to the action of
$K$, $L_K \phi^A = 0$, then the $1$-forms $T(K)$ and $K$
are proportional. As a consequence of Einstein's
equations, the Ricci $1$-form $R(K)$ (with components
$(R(K))_{\mu} \equiv R_{\mu \nu} K^{\nu}$) is then
proportional to $K$ as well,
\be
T(K) \, = \, -{\cal L} \, K \, \, ,
\; \; \; \; \;
R(K) \, = \, 8 \pi G \,
[{\cal L} \, - \,
\frac{1}{2} G_{AB} \, (d \phi^A | {\cal L}_{d \phi_B})] \, \, K \, .
\label{6}
\ee
Hence, in a stationary and axisymmetric spacetime,
the symmetry equations for $\phi^A$ together with
Einstein's equations immediately imply that the
{\em{Ricci circularity}} conditions,
\be
i_m \, \ast (k \wedge R(k)) \, = \, 0 \, ,
\; \; \; \; \;
i_k \, \ast (m \wedge R(m)) \, = 0 \, ,
\label{7}
\ee
are fulfilled.
Here we have used the operators $i_K$ and $\ast$ to
denote the inner product with respect to the vector
field $K$ and the Hodge dual, respectively. (Recall that
in $3+1$ dimensions we have
$\ast^2 \Omega = - (-1)^p \Omega$ and
$i_K \Omega = - \ast (K \we \ast \Omega)$ for an
arbitrary $p$-form $\Omega$.)

Next we apply the circularity theorem
\cite{Kund66}, \cite{Cart69},
implying that in an asymptotically flat,
stationary and axisymmetric spacetime,
the {\em{Frobenius integrability}} conditions
are satisfied,
\be
(m | \omega_k) \, = \, 0 \, ,
\; \; \; \; \;
(k | \omega_m) \, = \, 0 \, ,
\label{8}
\ee
provided that the Ricci circularity conditions (\ref{7})
hold (and vice versa). As usual, the twist $1$-form
$\omega_K$ assigned to an arbitrary $1$-form $K$
is defined as
\be
\omega_K \, \equiv \, \frac{1}{2} \, \ast (K \we dK) \, .
\label{9}
\ee

A simple proof of the circularity theorem is obtained
as follows: First, one notes that for an arbitrary Killing
field ($1$-form) $K$ the differential of $\omega_K$
fulfills the identity
\be
d \omega_K \, = \, \ast (K \we R(K)) \, .
\label{10}
\ee
Using the co-differential, $\cod = \ast d \ast$, and
the Laplacian, $- \Delta = \cod d + d \cod$,
eq. (\ref{10}) is obtained from the identities
$\cod K = 0$, $\cod (K \we dK) = - K \we \cod dK$ and
$- \Delta K = 2 R(K)$, which hold for arbitrary
{\em Killing} fields ($1$-forms) (see \cite{Heus93a}
for details). It is now easy to see that the Ricci
circularity conditions (\ref{7}) are the exterior
differentials of the integrability conditions (\ref{8}):
Clearly $[k,m] = 0$ implies
$L_m \omega_k = L_k \omega_m = 0$ (since the Hodge
dual commutes with the Lie derivative with respect
to a Killing field). Using $L_m = i_m d + d i_m$ and
the identity (\ref{10}), we immediately obtain
\be
d \, (m | \omega_k) \, = \, d i_m \, \omega_k \, =
\, -i_m d \, \omega_k \, = \, -i_m \ast (k \wedge R(k))
\label{11}
\ee
and, in the same manner, the corresponding expression
for $d(k | \omega_m)$. This demonstrates that integrability
implies Ricci circularity. The converse statement holds
as well: By vitue of eq. (\ref{11}), Ricci circularity implies
$d(m|\omega_k) = 0$. Hence the {\em function}
$(m|\omega_k)$ is constant. Since $m$ vanishes on
the rotation axis, $(m|\omega_k)$ vanishes identically
in every region of $(M,g)$ containing a part of the axis.
The fact that the same argument also holds for
$(k|\omega_m)$ concludes the proof.
In summary, we have the following result: \\

{\em{Let $(M,g)$ be an asymptotically flat,
stationary and axisymmetric spacetime
with Killing fields $k$ and $m$, and let $\phi$
denote a self-gravitating, minimally coupled mapping
from $(M,g)$ into a Riemannian manifold $(N,G)$,
$\phi$ being invariant under the action of $k$
and $m$. Then (in every connected domain of spacetime
containing a part of the rotation axis) the $2$-surfaces
orthogonal to $k$ and $m$ are integrable. }}

\section{The Field Equations}

In this section we give the basic
expressions for the Ricci tensor of a
spacetime manifold $(M,g)$ which admits
an {\em{integrable}} system of two
Killing fields, $k$ and $m$. Using Einstein's
equations for self-gravitating
harmonic mappings, these identities
reduce to the vacuum Ernst equations
for the metric functions of the $2$-dimensional
orbit manifold, plus an additional
set of equations involving the matter fields
and the metric of the manifold
orthogonal to the Killing orbits.

The integrability conditions imply
$M = \Sigma \times {\cal T}$ and $g = \sigma + \tau$,
where $(\Sigma,\sigma)$ and $({\cal T},\tau)$
are $2$-dimensional manifolds with
pseudo-Riemannian metric $\sigma$
and Riemannian metric $\tau$, respectively.
Parametrizing $\Sigma = {\cal R} \times SO(2)$
with $t$ and $\varphi$, we have
$k = \partial / \partial t$ and
$m = \partial / \partial \varphi$.
Choosing the co-ordinates $x^0 = t$,
$x^1 = \varphi$ $\in \Sigma$,
$x^2$, $x^3$  $\in {\cal T}$, and introducing
the adapted local basis of $1$-forms
\be
\theta^a \, = \, dx^a  , \; \; a \in \{ 0,1 \} \, ,
\; \; \; \; \; \;
\theta^i \, = \, dx^i  , \; \; i \in \{ 2,3 \} \, ,
\label{13}
\ee
the spacetime metric becomes
\be
g \, = \, \sigma_{ab} \, \theta^a \otimes \theta^b \; + \;
\tau_{ij} \, \theta^i \otimes \theta^j \, ,
\label{14}
\ee
where both $2$-dimensional metrics,
$\sigma$ and $\tau$, depend on
the co-ordinates $x^i$ of ${\cal T}$
only, $\sigma  =  \sigma(x^i)$,
$\tau = \tau (x^i)$, $i \in \{ 2,3 \}$.

The components of the Ricci tensor
with respect to the metric (\ref{14})
can be derived in an invariant manner.
However, before doing so, one has
to select the Killing field on the basis
of which one intends to formulate
the Ernst equations. The "good" (although
not the traditional) choice is to
consider the Killing field $m$ which
generates the axial symmetry \cite{Cart87}.
There are two reasons for this: First of
all, the norm of $k$ has no fixed sign if
spacetime is stationary, rather than
stationary in the {\em strict} sense.
As a consequence, the system of differential
equations formulated with respect
to $k$ turns out to be singular at the boundaries
of "ergoregions", appearing for
all rotating black hole solutions.
Secondly, if electromagnetic fields are also taken
into account, the Ernst equations can still be
derived from an action principle.
However, only the Ernst potentials which
are assigned to the axial Killing field $m$
lead to a {\em definite} action. The fixed
sign of the effective Lagrangian is,
however, a necessary requirement for
the applicability of the uniqueness proof
for rotating electrovac black holes
\cite{Mazu82}, \cite{Cart85}.

Parametrizing the metric $\sigma$ by the
three functions $\rho \equiv \sqrt{-det(\sigma)}$,
$X \equiv (m|m)$ and $A$, and introducing the
$2$-dimensional Riemannian metric
$\gamma = X \tau$ on ${\cal T}$,
\be
g \, = \, - \, \frac{\rho^2}{X} \, dt^2
\; + \; X \, (d\varphi \, + \, A \, dt )^2 \; + \;
\frac{1}{X} \, \gamma \, ,
\label{16}
\ee
one obtains the following set of differential identities
\be
\frac{1}{\rho} \, \codga (\rho \eform)
\, = \,
\frac{1}{X} \, [(\eform|\eform)^{(\gamma)} \, -
\, 2 \, R(m,m)] \, ,
\label{17}
\ee
\be
\frac{1}{\rho} \, \lapgamma \rho
\, = \,
- \frac{1}{X} \, tr^{(\sigma)} \, R \, ,
\label{18}
\ee
\be
\frac{1}{\rho} \, dA
\, = \,
2 \, \ast^{(\gamma)} (\frac{\omega}{X^2}) \, ,
\label{19}
\ee
\be
R_{ij} \, + \, \gamma_{ij} \frac{R(m,m)}{X^2}
\, = \,
\kappa^{(\gamma)} \, \gamma_{ij} \, - \,
\frac{1}{\rho}  \nabiga \nabjga \rho \, - \,
\frac{\eform_i \bar{\eform}_j \, + \, \eform_j
\bar{\eform}_i}{4 \, X^2} \, ,
\label{20}
\ee
where $\codga$, $\lapgamma$, $\ast^{(\gamma)}$ and
$\kappa^{(\gamma)}$ denote the co-differential,
the Laplacian, the Hodge dual and the Gauss curvature
with respect to the $2$-dimensional Riemannian
metric $\gamma$, and
$tr^{(\sigma)} R \equiv \sigma^{ab} R_{ab}$.
The complex $1$-form
${\cal E}$ is defined as
\be
{\cal E} \, = \, - \, dX \, + \, 2 i \, \omega \, ,
\label{21}
\ee
and the twist form $\omega \equiv \omega_m$
associated to $m$ fulfills
the general identity (\ref{10}),
\be
d \omega \, = \, \ast (m \we R(m)) \, .
\label{22}
\ee
The formulas (\ref{17})-(\ref{22}) are
geometrical identities. They are valid if
the integrability conditions (\ref{8})
are fulfilled. Einstein's equations imply
that this is the case for vacuum models,
electromagnetic fields and, as we
have demonstrated in the previous
section, for invariant scalar mappings.

It is well-known that the above formulas experience
a significant simplification for both vacuum and
electrovac spacetimes. This is due to the
fact that the r.h.s. of eq. (\ref{18}) vanishes in both
situations. This is also the case for arbitrary {\em{harmonic}} mappings,
to which we shall restrict our attention
in the remainder of this paper. Hence
we consider
\be
{\cal L} \, = \, \frac{1}{2} \,
\parallel d \phi \parallel ^2 \, ,
\; \; \; \; \;
T \, = \, G_{AB} \, d \phi^A \otimes \phi^B \, - \,
\frac{1}{2} \, \parallel d \phi \parallel ^2 \, g \, ,
\label{23}
\ee
and use Einstein's equations in the above
identities. As is immediately observed
from eq. (\ref{6}), the invariance properties,
$L_k \phi^A = L_m \phi^A = 0$,
imply that the Ricci $1$-forms with
respect to both Killing fields vanish,
\be
R(k) \, = \, 0 \, ,
\; \; \; \; \;
R(m) \, = \, 0 \, ,
\; \; \; \; \;
tr^{(\sigma)} \, R \, = \, 0 \, .
\label{24}
\ee
In addition, the components of $R_{\mu \nu}$
on $({\cal T}, \gamma)$ become
\be
R_{ij} \, = \, 8 \pi G \; G_{AB} \, \phi^A,_i \phi^B,_j \, .
\label{25}
\ee
Since $R(m)$ vanishes, we conclude from
eq. (\ref{22}) that the twist form is
closed. Hence, there locally exists a twist
potential $Y$, such that $\omega = dY$.
Since the complex $1$-form ${\cal E}$ is
then closed as well, we can introduce the
same Ernst potential as in the vacuum
case \cite{Ernst68a}, \cite{Ernst68b},
\be
E \, = \, - \, X \, + \, i \, Y \, ,
\; \; \; \mbox{where} \; \; dY \, = \; \omega \, .
\label{26}
\ee
Using the complex potential $E$ and
eqs. (\ref{24}), (\ref{25}) for the
Ricci tensor, the field equations are
now obtained from the identities
(\ref{17})-(\ref{20}),
\be
\lapgamma E \, + \,
\frac{1}{\rho} \, (d \rho | dE)^{(\gamma)} \, + \,
\frac{1}{X} \, [(dE|dE)^{(\gamma)} \, = \, 0 \, ,
\label{17a}
\ee
\be
\lapgamma \rho \, = \, 0 \, ,
\label{18a}
\ee
\be
\frac{1}{\rho} \, dA
\, = \,
2 \, \ast^{(\gamma)} (\frac{dY}{X^2}) \, ,
\label{19a}
\ee
\be
\kappa^{(\gamma)} \, \gamma_{ij} \, - \,
\frac{1}{\rho}  \nabiga \nabjga \rho \, = \,
\frac{E,_i \bar{E},_j \, + \, E,_j \bar{E},_i}{4 \, X^2} \, + \,
8 \pi G \; G_{AB} \, \phi^A,_i \phi^B,_j \, ,
\label{20a}
\ee
where $X = - \Re e(E)$, $Y = \Im m(E)$.
(Note that we have used the general identity
$\cod(f \, dg) = \ast (df \we \ast dg) +
f \cod dg = - [(df | dg) + f \Delta g]$ for
arbitrary functions $f$ and $g$.) In addition
to the above formulas one also has the
matter equations, $\Delta \phi^A +
\Gamma^A_{\; BC} (\phi) (d \phi^A | d \phi^B) = 0$,
which are obtained by varying the action
(\ref{1}) with Lagrangian  (\ref{23}) with
respect to $\phi^A$. (Here $\Gamma^A_{\; BC}$
denote the Christoffel symbols
assigned to the metric $G$ of the target
manifold $(N,G)$, see \cite{Misn78} or
\cite{Eell78}, \cite{Eell88} for details).
Since the $\phi^A$ do not depend on $t$
and $\varphi$, we have $(d \phi^A | d \phi^B) =
X (d \phi^A | d \phi^B)^{(\gamma)}$
and $\Delta \phi^A = X \, [\lapgamma \phi^A +
(d \, ln \rho|d \phi^A)^{(\gamma)}]$.
Hence the matter equations assume the form
\be
\lapgamma \phi^A \, + \, \frac{1}{\rho}
(d \rho|d \phi^A)^{(\gamma)} \, + \,
\Gamma^A_{\; BC}
\, (d \phi^A | d \phi^B)^{(\gamma)} \, = \, 0 \, .
\label{27}
\ee

The formulas (\ref{17a})-(\ref{27}) represent
the complete set of field equations
for the functions $E$, $\rho$, $\phi^A$ and
the $2$-dimensional Riemannian metric
$\gamma$. All quantities depend only
on the co-ordinates $x^2$ and $x^3$.
The crucial observation consists in the fact
that the matter fields do {\em not}
appear in the Einstein equations
(\ref{17a})-(\ref{19a}) which completely
determine the metric $\sigma$.
As in the vacuum case (and for the
Einstein-Maxwell system),
$\rho = \sqrt{-det(\sigma)}$ is a
harmonic function on
$({\cal T},\gamma)$. This yields a
further, considerable
simplification: Using asymptotic
flatness, ordinary Morse
theory \cite{Mors45} implies that
the number of local maxima
plus the number of local minima
minus the number of saddle
points of $\rho$ is a topological
invariant, which vanishes
if the domain is simply connected
and all critical points are
non-degenerate. Together with the
maximum principle,
this theorem immediately implies
that $\rho$ has no
non-degenerate inner critical points
at all. Using the special
Morse theory for harmonic functions one can, in addition,
exclude degenerate critical points as well.
Hence, if $\lapgamma \rho = 0$, it is possible to choose
$\rho$ as one of the co-ordinates on $({\cal T},\gamma)$
\cite{Cart87}, \cite{Cart73}. Denoting the remaining
co-ordinate with $z$, one can introduce a conformal factor,
$exp(2h(\rho,z))$, such that the metric $\gamma$ assumes
the diagonal form $\gamma =  e^{2 \, h} ( d\rho^2 + dz^2 )$.
Thus, the spacetime metric is finally parametrized by only
three functions $X$, $A$ and $h$ of two variables $\rho$
and $z$ \cite{Papa53},
\be
g \, = \, - \, \frac{\rho^2}{X} \, dt^2
\; + \; X \, (d\varphi \, + \, A \, dt )^2 \; + \; \frac{1}{X} \,
e^{2 \, h} \, (d \rho^2 \, + \, dz^2) \, .
\label{28}
\ee

It remains to express equations (\ref{20a}) in terms
of the co-ordinates $\rho$ and $z$. The second covariant
derivatives of $\rho$ transform into ordinary first
derivatives of the conformal factor $e^{2h}$ with respect
to $\rho$ and $z$. Using $\Gamma^{\rho}_{\, \rho \rho} =
\Gamma^{z}_{\, z \rho} = -\Gamma^{\rho}_{\, z z} = h,_{\rho}$
and $\Gamma^{z}_{\, z z} = \Gamma^{\rho}_{\, \rho z} =
-\Gamma^{z}_{\, \rho \rho} = h,_z$, one immediately finds
\be
\nabzga \nabzga \rho \, = \,
- \, \nabrga \nabrga \rho \, = \, h,_{\rho} \, ,
\; \; \; \;
\nabrga \nabzga \rho \, = \,
\nabzga \nabrga \rho \, = \, h,_z \, ,
\label{eq:kov2}
\ee
\be
R^{(\gamma)}_{ij} \, = \, \kappa ^{(\gamma)} \, \gamma_{ij} \, ,
\; \; \; \;
\kappa ^{(\gamma)} \, = \, - \, \lapgamma h \, = \,
- \, e^{- 2 \, h} \, \lapdelta h \, ,
\label{eq:kappadef}
\ee
where $\lapdelta$ denotes the flat Laplacian with respect to the
co-ordinates $\rho$ and $z$. Hence, in terms of Weyl co-ordinates,
equation (\ref{20a}) assumes the form
\bea
\frac{1}{\rho} \, h,_{\rho}
& = &
\frac{1}{2} (R_{\rho \rho} - R_{z z}) \, + \frac{1}{4 X^2} \,
(E,_{\rho} \Ebar,_{\rho} \, - \, E,_z \Ebar,_z) \, ,
\label{30a} \\
\frac{1}{\rho} \, h,_z
& = &
\frac{1}{2} (R_{\rho z} + R_{z \rho}) \, + \frac{1}{4 X^2} \,
(E,_{\rho} \Ebar,_z \, + \, E,_z \Ebar,_{\rho}) \, ,
\label{30b} \\
- \, \lapdelta \, h
& = &
\frac{1}{2} (R_{\rho \rho} + R_{z z}) \, + \frac{1}{4 X^2} \,
(E,_{\rho} \Ebar,_{\rho} \, + \, E,_z \Ebar,_z) \, ,
\label{30c}
\eea
where $R_{ij}$ is given by eq. (\ref{25}), and $i,j \in \{\rho, z \}$.

The linearity of the above equations in the metric function $h$
suggests the partition
\be
h \, = \, \hvac \, + \, 8 \pi G \; \hphi \, ,
\label{31}
\ee
where $\hvac$ is required to fulfill
eqs. (\ref{30a})-(\ref{30c}) with $R_{ij} = 0$.
The entire set of field equations is now solved as follows: \\

\noindent
(i) {\ul{\em{Vacuum equations:}}}
Like in the vacuum case, one first solves the $2$-dimensional
boundary value problem in the
$(\rho, z)$ plane (with fixed, flat background metric $\delta$)
for the Ernst potential $E$,
being subject to the Ernst equation
\be
\frac{1}{\rho} \, \nabdel(\rho \, \nabdel \, E) \,
+ \, \frac{(\nabdel E|\nabdel E)}{X} \, = \, 0 \, .
\label{VAC1}
\ee
Subsequently, one obtains the metric functions $A$ and $\hvac$
by quadrature,
\be
\frac{1}{\rho} \, A,_{\rho} \, = \, \frac{1}{X^2} \, Y,_z \, ,
\; \; \; \; \;
\frac{1}{\rho} \, A,_z \, = \, - \, \frac{1}{X^2} \, Y,_{\rho} \, ,
\label{VAC2}
\ee
\be
\frac{1}{\rho} \, \hvac,_{\rho} \, = \, \frac{1}{4 X^2} \,
[E,_{\rho} \Ebar,_{\rho} \, - \, E,_z \Ebar,_z] \, ,
\; \; \; \; \;
\frac{1}{\rho} \, \hvac,_z \, = \, \frac{1}{4 X^2} \,
[E,_{}\rho \Ebar,_z \, + \, E,_z \Ebar,_{\rho}] \, ,
\label{VAC3}
\ee
where $X = - \Re e(E)$, $Y = \Im m(E)$. \\

\noindent
(ii) {\ul{\em{Matter equations:}}}
Like the Ernst equation,
the field equations for the matter fields $\phi^A$,
\be
\frac{1}{\rho} \, \nabdel(\rho \, \nabdel \, \phi^A) \,
+ \, \Gamma^A_{\; BC} \; (\nabdel \phi^B | \nabdel \phi^C) \, = \, 0 \, ,
\label{MAT1}
\ee
involve no unknown metric functions.
Having solved the boundary value problem
for $\phi^A (z, \rho)$, the remaining metric function
$\hphi$ is also obtained by quadrature,
\be
\frac{1}{\rho} \, \hphi,_{\rho} \, = \, \frac{G_{AB}}{2} \,
[\phi^A,_{\rho} \phi^B,_{\rho} \, - \,
\phi^A,_z \phi^B,_z] \, ,
\; \; \; \; \;
\frac{1}{\rho} \, \hphi,_z \, = \, \frac{G_{AB}}{2} \,
[\phi^A,_{\rho} \phi^B,_z \, + \,
\phi^A,_z \phi^B,_{\rho}] \, .
\label{MAT2}
\ee

\section{The Effective Action}

The uniqueness proofs of Bunting \cite{Bunt83}
and Mazur \cite{Mazu82} for the Kerr-Newman
metric rely on the circumstance that the field
equations can be obtained from an effective,
definite action which is defined on the $2$-dimensional
Riemannian manifold ${\cal T}$. In this section we
establish the corresponding result for the equations
describing stationary and axisymmetric selfgravitating
harmonic mappings. We shall demonstrate that both, the
Ernst equation (\ref{VAC1}) for the potential $E$ and the
matter equations (\ref{MAT1}) for the fields $\phi^A$, can
be gained from the effective Lagrangian ${\cal L}_{eff} =
\rho \sqrt{\gamma} \kappa^{(\gamma)}$, $\kappa^{(\gamma)}$
denoting the Gauss curvature of $({\cal T},\gamma)$.

Before considering the variational principle, let us first
recall the fact that the Ernst equation (\ref{VAC1})
coincides with the integrability condition for the system
(\ref{VAC3}) and, in addition, guarantees the consistency
of these equations with the Poisson equation
for the function $\hvac$,
\be
- \, \lapdelta \, \hvac \, = \, \frac{1}{4 X^2} \,
(E,_{\rho} \Ebar,_{\rho} \, + \, E,_z \Ebar,_z) \, .
\label{lapvac}
\ee
In exactly the same manner,
the matter equations (\ref{MAT1}) are identical with the
integrability conditions for the system (\ref{MAT2}) and
guarantee their consistency with the second order equation
\be
- \, \lapdelta \, \hphi \, = \,
\frac{G_{AB}}{2} \,
[\phi^A,_{\rho} \phi^B,_{\rho} \, + \,
\phi^A,_z \phi^B,_z]
\label{lapphi}
\ee
for the function $\hphi$. Let us, as an example, demonstrate the consistency
of eqs. (\ref{MAT2}) and (\ref{lapphi}). Differentiating
$\hphi,_{\rho}$ with respect to ${\rho}$ and
$\hphi,_z$ with respect to $z$ gives
\bea
\hphi,_{\rho \rho} + \hphi,_{z z} & = &
\rho \, \phi^A,_{\rho} \{
G_{AB} (\phi^B,_{z z} + \phi^B,_{\rho \rho} +
\frac{1}{\rho} \phi^B,_{\rho}) +
(G_{AB,C} - \frac{1}{2} G_{BC,A}) \, \phi^B,_{z} \phi^C,_{z}
\nonumber  \\ & + &
\frac{1}{2} G_{AB,C} \, \phi^B,_{\rho} \phi^C,_{\rho} \} -
\frac{G_{AB}}{2} \, ( \phi^A,_{\rho} \phi^B,_{\rho}
+ \phi^A,_{z} \phi^B,_{z} )
\, = \, \lapdelta \hphi \, ,
\nonumber
\eea
where we have used the matter eq.
(\ref{MAT1}) and the identities
$\Gamma_{BC|A} \phi^A,_{\rho} \phi^B,_{\rho} \phi^C,_{\rho}$ $=$
$\frac{1}{2} G_{AB,C} \phi^A,_{\rho} \phi^B,_{\rho} \phi^C,_{\rho}$
and
$\Gamma_{BC|A} \phi^A,_{\rho} \phi^B,_{z} \phi^C,_{z}$ $=$
$(G_{AB,C} - \frac{1}{2} G_{BC,A}) \phi^A,_{\rho} \phi^B,_{z} \phi^C,_{z}$.

Let us now consider the reduction of the
matter action with respect to the metric (\ref{14}).
Using the fact that the matter fields do not depend on the
co-ordinates of $(\Sigma , \sigma)$, we obtain
\bdm
S^{(\phi)} = \,
\frac{1}{2} \int_{M}
\parallel d \phi \parallel ^2 \, \eta \, = \,
\frac{1}{2} \int_{M}
G_{AB} (d \phi^A | d \phi^B)^{(\gamma)} \sqrt{\gamma} \sqrt{|\sigma|} \,
dx^0 dx^1 d \rho \, dz
\edm
\be
= \, Vol(\Sigma) \, \frac{1}{2} \int_{{\cal T}}
G_{AB} (\nabdel \phi^A | \nabdel \phi^B)^{(\delta)}
\, \rho \, d \rho \, dz \, .
\label{matter}
\ee
Together with eq. (\ref{lapphi}), this suggests
that $- \rho \lapdelta \hphi$ is the effective
Lagrangian for equations (\ref{MAT1}).
As a matter of fact, it is not hard to verify
that the effective Lagrangian
for the Ernst equation (\ref{VAC1})
{\em and} the matter equations (\ref{MAT1}) is
proportional to the Gauss curvature of $({\cal T}, \gamma)$,
\be
{\cal L}_{eff} \, = \,
- \rho \, \lapdelta (\hphi + 8 \pi G \, \hvac) \, = \,
- \rho \, \sqrt{\gamma} \, \lapgamma h \, = \,
\rho \, \sqrt{\gamma} \, \kappa ^{(\gamma)} \, ,
\label{leff}
\ee
where $\lapdelta h$ is obtained from eqs.
(\ref{lapvac}) and (\ref{lapphi}). Thus, we have the following result: \\

{\em{Let $(M,g)$ be an asymptotically flat,
stationary and axisymmetric spacetime
with Killing fields $k$ and $m$.
Let $\phi$ denote a mapping with harmonic action from $(M,g)$
into a Riemannian manifold $(N,G)$, $\phi$ being invariant
under the action of $k$ and $m$.
Then the metric and the fields $\phi$ are obtained
by solving the vacuum Ernst equation (\ref{VAC1}) and the matter
equations (\ref{MAT1}), which are the
Euler-Lagrange equations for the effective Lagrangian}}
\be
{\cal L}_{eff} \, = \, \rho \,
[\frac{(\nabdel E | \nabdel \Ebar)}{(E + \Ebar)^2} \; + \; 8 \pi G \;
\frac{G_{AB} \, (\nabdel \phi^A | \nabdel \phi^B)}{2}] \, .
\label{efflagr}
\ee

As usual, it is convenient to introduce prolate spheroidal
co-ordinates $x$ and $y$ (see e.g. \cite{Cart73})
in order to distinguish between the
horizon and the rotation axis,
\be
\rho^2 \; = \; \mu^2 \; (x^2 \, - \, 1) \, (1 \, - \, y^2) \; ,
\; \; \; \; \; \;
z \; = \; \mu \; x \, y \; ,
\label{prolate}
\ee
where $\mu$ is an arbitrary positive constant.
Note that $(\rho,z) \rightarrow (x,y)$ maps the upper half
plane to the semi-strip
${\cal S} = \{(x,y) \, | \, x \geq 1\, , |y| \leq 1 \}$,
where the boundary $\rho = 0$ splits into three parts,
being the horizon, $x = 0$, and the northern and southern
segments of the rotation axis, $y=1$ and $y=-1$, respectively.
In terms of prolate spheroidal co-ordinates the $2$-dimensional
metric $\gamma$ becomes
\be
\gamma \, = \, e^{2 \, h} \, (d\rho^2 \, + \, dz^2) \, = \,
e^{2 \, h} \, \mu^2 \, (x^2 - y^2) \,
(\frac{dx^2}{x^2 - 1} \, + \, \frac{dy^2}{1 - y^2}) \, .
\label{gaxy}
\ee
In order to solve the Ernst equation, it is very helpful
to introduce the potential $\eps$,
\be
\eps \, = \, \frac{1 \, + \, E}{1 \, - \, E} \, ,
\label{eq:epsilonernst}
\ee
which parametrizes the points in the semi-plane $X \geq 0$
by points in the unit disc
${\cal D} = \{\eps \, | \, \eps \overline{\eps} \leq 1 \}$, since
\be
X \, = \, \frac{1 - \eebar}{|1 + \eps|^2} \, ,
\; \; \; \; \;
Y \, = \, \frac{i \, (\epsbar - \eps)}{|1 + \eps|^2} \, .
\label{XYeps}
\ee
In terms of prolate spheroidal co-ordinates,
the effective Lagrangian for $\eps (x,y)$,
describing the mapping from the semi-strip ${\cal S}$
into the unit disc ${\cal D}$, becomes
\be
{\cal L}_{eff}^{(vac)} \, = \, \mu \,
\frac{(x^2 -1) \eps,_x \epsbar,_x \,
+ \, (1 - y^2) \eps,_y \epsbar,_y}{(1 \, - \, \eps \epsbar)^2} \, .
\label{actdef}
\ee

Before we continue, let us, as an example,
consider mappings
$\phi$ from spacetime into the pseudo-sphere $N = PS^2$,
\be
\phi \, = \, (\chi \, , \varphi) \, ,
\; \; \; \; \;
G_{AB} (\phi) \, = \, diag \, (1 \, , sh^2 \chi) \, .
\label{NPS}
\ee
Parametrizing $PS^2$ with co-ordinates
$w = \tanh  (\frac{\chi}{2}) e^{i \varphi}$
in the unit disc
${\cal D} = \{w \, | \, w \overline{w} \leq 1 \}$,
we obtain
\be
G_{AB} (\phi) \, (d \phi^A | d \phi^B) \, = \,
d \chi^2 \, + sh^2 \chi \, d \varphi^2 \, = \,
4 \, \frac{(dw|d \overline{w})}{(1 \, - \, w \overline{w})^2} \, ,
\label{ww}
\ee
which shows that the effective matter Lagrangian in this case
becomes
\be
{\cal L}^{(\phi)}_{eff} \, = \, 16 \pi \, G \, \mu \,
\frac{(x^2 -1) w,_x w,_x \, + \,
(1 - y^2) w,_y w,_y}
{(1 \, - \, w \overline{w})^2} \, .
\label{mattww}
\ee
This demonstrates that both $\eps$ and $w$
satisfy exactly the same equation, if $\phi$
describes a stationary and axisymmetric mapping
from spacetime into the pseudo-sphere.

\section{Uniqueness of the Ernst potential}

In this section we briefly recall the derivation
of the Kerr solution and the arguments establishing
its uniqueness. Applied to the situation under
consideration, these arguments guarantee the uniqueness
of the metric $\sigma$, (i.e. of the functions
$X$ and $A$) and the function $\hvac$.
This clearly implies that the only possible
deviation from the Kerr solution can
occur via $\hphi$.
Hence, in order to establish the uniqueness theorem,
it will remain to prove that $\hphi$ vanishes
identically. Once the circularity theorem is established,
this is in fact the only additional step which has to be
performed in order to extend the vacuum no-hair
theorem for the Kerr solution to selfgravitating
harmonic mappings. In the next section
we shall prove that $\hphi$ vanishes as a
consequence of asymptotic flatness.

It is well known that a linear Ansatz shows that
the Ernst equation for the Lagrangian (\ref{actdef})
admits the simple solution
\be
\eps \, = \, p x \, + \, i \, q y \, ,
\label{sol1}
\ee
provided that the real constants $p$ and $q$ are subject to the
condition $p^2 + q^2 = 1$.
Inserting the solution (\ref{sol1})
into the expressions (\ref{XYeps}) for $X$ and $Y$,
and integrating subsequently eqs. (\ref{VAC2}) and (\ref{VAC3})
for $A$ and $\hvac$ in prolate spheroidal co-ordinates,
yields the metric functions
\be
X \, = \, \frac{1 - (px)^2 - (qy)^2}{(1 + px)^2 + (qy)^2} \, ,
\; \; \; \;
Y \, = \, \frac{2 \, q y}{(1 + px)^2 + (qy)^2} \, ,
\label{XY}
\ee
\be
A \, = \, 2 \mu \,(q / p) \,
\frac{(1 - y^2) \, (1 + px)}{1 - (px)^2 - (qy)^2} \, ,
\; \; \; \;
e^{2h} \, = \,
\frac{1 - (px)^2 - (qy)^2}{p^2 \, (x^2 - y^2)} \, .
\label{Ah}
\ee
It is immediately observed that
the norm $X$ of the Killing filed $m$
does neither vanish on the rotation axis,
nor is it definite in the entire semi-strip ${\cal S}$.
Hence, the above solution is physically not acceptable.
In order to obtain an
asymptotically flat solution with $X \geq 0$,
it remains to construct the
{\em{conjugate}} solution
$(\Xhat, \Ahat, \hhat)$ \cite{Chan83}.
The latter is obtained from the observation that the metric
$\sigma$ is invariant under the $(t,\varphi)$-rotation
\be
t \, \rightarrow \, \that \, = \, \varphi \, ,
\; \; \; \; \;
\varphi \, \rightarrow \, \phihat \, = \, - \, t \, ,
\label{rotplane}
\ee
provided that $X$, $A$ and $\hvac$ are transformed according to
\be
\Xhat \, = \, X \, [A^2 - \frac{\rho^2}{X^2}] \, ,
\; \; \; \;
\Ahat \, = \, -\, A \, [A^2 - \frac{\rho^2}{X^2}]^{-1} \, ,
\; \; \; \;
e^{2 \hhatvac} \, = \, e^{2 \hvac} \, [A^2 - \frac{\rho^2}{X^2}] \, .
\label{consol}
\ee
Performing these transformations,
one obtains after some straightforward
algebraic manipulations the Kerr solution
in prolate spheroidal coordinates:
\be
\Xhat \, = \, \frac{\mu^2}{p^2} \, (1 - y^2) \,
\frac{[(1 + px)^2 + q^2]^2 \, - \,
p^2 q^2 (x^2-1)(1-y^2)}{(1 + px)^2 + (qy)^2} \, ,
\label{KerrX}
\ee
\be
\Ahat \, = \, -2 \, \frac{q p}{\mu} \,
\frac{(1 + px)}{[(1 + px)^2 + q^2]^2 \, - \,
p^2 q^2 (x^2-1)(1-y^2)} \, ,
\label{KerrA}
\ee
\be
e^{2 \hhatvac} \, = \, \Xhat \;
\frac{(1 + px)^2 + (qy)^2}{p^2 \, (x^2 - y^2)} \, .
\label{Kerrh}
\ee
In addition, the rotation potential
$\Yhat$ ($d \Yhat = 2 \hat{\omega}$) is
computed from $\Ahat$ using the equations
$\dx \Yhat = - [\mu (x^2-1)]^{-1} \Xhat^2 \dy \Ahat$ and
$\dy \Yhat = [\mu (1-y^2)]^{-1} \Xhat^2 \dx \Ahat$, which yield
\be
\Yhat \, = \, 2 \, \frac{q \mu^2}{p^2} \, y \,
[(3-y^2) \, + \, \frac{q^2 \, (1 - y^2)^2}{(1 + px)^2 + (qy)^2}] \, .
\label{KerrY}
\ee
The spacetime metric in prolate spheroidal
coordinates eventually becomes
\be
g \, = \, - \Vhat dt^2 + 2 \What dt d\varphi + \Xhat d \varphi^2
+ \frac{e^{2 \hhatvac}}{\Xhat}
\, e^{8 \pi G \, 2 \hphi}
\, \mu^2 (x^2 - y^2) (\frac{dx^2}{x^2-1} +
\frac{dy^2}{1-y^2}) \, ,
\label{Kerrds}
\ee
where $\What = \Ahat \Xhat = - A X$ and $\Vhat = -X$.
The coupling to the matter fields $\phi^A$
enters via the function $\hphi$, being a solution of eqs.
(\ref{MAT1}), (\ref{MAT2}).

It is immediately verified from
eqs. (\ref{KerrX}) and (\ref{KerrY}) that
the real and imaginary part, $-\Xhat$ and $\Yhat$,
of the Ernst potential satisfy the
following boundary conditions
with respect to prolate spheroidal co-ordinates
\cite{Cart87}:
As $x \rightarrow \infty$,
\bea
\Xhat & = &
(1 - y^2) \,[\mu^2 \, x^2 \, + \, {\cal O}(x)] \, ,
\label{Xasy} \\
\Yhat & = &
2 \, J \, y \, (3 - y^2) \, + \, {\cal O}(x^{-1}) \, ,
\label{Yasy}
\eea
where we have used $J = ma$,
$m = \mu/p$, $a =\mu q/p$.
In the vicinity of the rotation axis,
$y \rightarrow \pm 1$, we have
\be
\Xhat \, = \, {\cal O}(1 - y^2) \; , \;
\; \; \; \;
(1 - y^2) \, \frac{\Xhat,_y}{\Xhat} \, = \,
\mp 2 \, + {\cal O}(1 - y^2) \; ,
\label{Xaxis}
\ee
\be
\Yhat,_x \, = \, {\cal O}(1 - y^2) \; , \;
\; \; \; \;
\Yhat,_y \, = \, {\cal O}(1 - y^2) \; ,
\label{Yaxis}
\ee
which shows that $\partial_y \Xhat / \Xhat$ remains finite,
although $\Xhat$ vanishes on the rotation axis.
Note that on the horizon, i.e. for $x = 1$, both functions
$\Xhat$ and $\Yhat$ behave perfectly regular,
\be
\Xhat \, = \, {\cal O}(1) \; , \;
\; \; \; \;
\Xhat ^{-1} \, = \, {\cal O}(1) \, ,
\label{Xhori}
\ee
\be
\Yhat,_x \, = \, {\cal O}(1) \; , \;
\; \; \; \;
\Yhat,_y \, = \, {\cal O}(1) \, .
\label{Yhori}
\ee

The metric (\ref{Kerrds}) with $h^{(\phi)} \equiv 0$ was
first derived by Kerr in 1963
\cite{Kerr63}, \cite{Kerr65}. It represented the first known
asymptotically flat exact solution of a rotating source in
general relativity. The $2$-parameter family is characterized
by the total mass $m$ and the total angular momentum $J$.
On the basis of the strong rigidity theorem, various
authors have contributed to the proof that the Kerr
metric describes the only asymptotically flat, stationary
vacuum black hole solution (see the introduction for references).
In the presence of electromagnetic fields,
an equivalent theorem applies to the $3$-parameter
Kerr-Newman family.

Later, Bunting \cite{Bunt83} and Mazur \cite{Mazu82}
succeeded in improving the classical arguments
leading to the uniqueness theorem for
the Kerr (-Newman) family. The reasoning relies on the regularity
and ellipticity of the $2$-dimensional
boundary value problem described by the Lagrangian (\ref{actdef})
and the boundary conditions (\ref{Xasy})-(\ref{Yhori}).
The basic observation consists in the fact that the vacuum
and electrovac Ernst equations are non-linear sigma-model
equations with symmetric spaces
$SU(1,1)/U(1)$ and $SU(1,2)/S(U(1) \times U(1))$, respectively.
In the vacuum case this is seen from the effective
Lagrangian (\ref{actdef}), which obviously
describes a harmonic map from the
semi-strip ${\cal S}$
(with metric $diag ((x^2-1)^{-1},(1-y^2)^{-1})$)
into the unit disc ${\cal D}$, being (as a symmetric space)
equivalent to $SU(1,1)/U(1)$.
Within this framework,
Robinson's identity \cite{Robi75}
is recovered as a special case of a divergence
relation for non-linear sigma-models
\cite{Mazu82}. The identity implies that
two solutions of the Ernst equations,
$(\hat{X}_1, \hat{Y}_1)$ and $(\hat{X}_2, \hat{Y}_2)$,
are identical if they are subject
to the same boundary conditions
(\ref{Xasy})-(\ref{Yhori}).

Applied to the problem under consideration,
the uniqueness theorem for the vacuum Kerr
metric guarantees that the metric
(\ref{Kerrds}), with the functions
$\hat{X}$, $\hat{A}$ and $\hat{h}^{(vac)}$
according to eqs. (\ref{KerrX})-(\ref{Kerrh}),
is the unique asymptotically flat solution
of the Einstein equations with harmonic fields.
Hence, in order to establish the uniqueness theorem
for this system, it remains to prove that the
additional metric function $h^{(\phi)}$ and the matter fields
$\phi^A$ are uniquely determined by eqs. (\ref{MAT1}),
(\ref{MAT2}), asymptotic flatness
and the boundary conditions for $\phi^A$.

\section{Uniqueness of the Matter Solution}

Let us now show that the matter equations (\ref{MAT1}),
(\ref{MAT2}) admit only the trivial solution
$\phi^A =$ constant, $h^{(\phi)} = 0$.
This is a consequence of
asymptotic flatness, the (weak) asymptotic fall-off
conditions
\be
\phi^A \, = \, \phi^A_{\infty} \, + \, {\cal O}(r^{-1}) \, ,
\; \; \; \; \;
\phi^A,_{\vartheta} \, = \, {\cal O}(r^{-1}) \, ,
\; \; \; \; \;
\phi^A,_{r} \, = \, {\cal O}(r^{-2}) \, ,
\label{ac1}
\ee
and the requirement that $G_{AB}(\phi)$ and the derivatives
of the scalar fields with respect to $r$ and $\vartheta$
remain finite at the boundary of the domain of outer communications.
As usual, $r$ and $\vartheta$ denote Boyer-Lindquist
co-ordinates \cite{BL},
\be
r \, = \, m \, (1 \, + \, p \, x) \, ,
\; \; \; \; \;
\cos \vartheta \, = \, y \, ,
\label{bl}
\ee
in terms of which we have
${\cal S} = \{ (r,\vartheta) \, | \, r \geq r_h , \,
\vartheta \in [0, \pi] \}$
(where $r = r_h$ denotes the event horizon).

To start, we note that eq. (\ref{MAT2})
for $h^{(\phi)}$ in prolate spheroidal co-ordinates
becomes
\be
h^{(\phi)},_x \, = \,
\frac{G_{AB}}{2} \frac{1 - y^2}{x^2 - y^2} \,
[x (x^2 - 1) \phi^A,_x \phi^B,_x -
x (1 - y^2) \phi^A,_y \phi^B,_y -
2y (x^2 - 1) \phi^A,_x \phi^B,_y] \, ,
\label{hx}
\ee
\be
h^{(\phi)},_y \, = \,
\frac{G_{AB}}{2} \frac{x^2 - 1}{x^2 - y^2} \,
[y (x^2 - 1) \phi^A,_x \phi^B,_x -
y (1 - y^2) \phi^A,_y \phi^B,_y +
2x (1 - y^2) \phi^A,_x \phi^B,_y] \, ,
\label{hy}
\ee
from which we also obtain the asymptotic behavior of
the derivative of $h^{(\phi)}$ with respect to $r$,
\bea
h^{(\phi)},_r & = &
\frac{G_{AB}}{2} \, \{ \, \sin^2 \vartheta
( \, [r + {\cal O}(1)] \, \phi^A,_r \phi^B,_r \, - \,
[r^{-1} + {\cal O}(r^{-2})] \,
\phi^A,_{\vartheta} \phi^B,_{\vartheta} )
\nonumber \\
& + &
\sin(2 \vartheta)
( \, [1 + {\cal O}(r^{-1})] \, \phi^A,_r \phi^B,_{\vartheta} ) \, \} \, .
\label{hr}
\eea
By virtue of the fall-off conditionsas (\ref{ac1})
for the scalar fields, all terms
in the bracket on the r.h.s. of eq.
(\ref{hr}) are of ${\cal O}(r^{-3})$.
Hence, the metric function $h^{(\phi)}(r, \vartheta)$
has the asymptotic property
\be
\lim_{r \rightarrow \infty} r^2 \, h^{(\phi)},_r \, = \, 0 \, .
\label{r2h}
\ee
In a similar way one establishes
$\lim_{r \rightarrow \infty} r h^{(\phi)},_{\vartheta} = 0$.
(Note also that asymptotic flatness requires
that $exp(2h)$ has an asymptotic expansion in terms
of $1/r$. Equation (\ref{r2h}) then shows that
$exp(8 \pi G \, 2h^{(\phi)}) = 1 + {\cal O}(r^{-2})$,
from which we find, using the vacuum solution
(\ref{Kerrh}) in Boyer-Lindquist coordinates,
\be
\frac{e^{2 \hhatvac}}{\Xhat}
\, e^{8 \pi G \, 2 \hphi} \, = \,
\frac{r^2 + a^2 \cos^2 \vartheta}{r^2
- 2mr + m^2 \sin^2 \vartheta +
a^2 \cos^2 \vartheta} [1 + {\cal O}(r^{-2})] \, = \,
1 + \frac{2m}{r} + {\cal O}(r^{-2}) \, ,
\label{asexp}
\ee
as $r \rightarrow \infty$.
This shows that the the first {\em two}
terms in the asymptotic expansion of the
metric coefficient of $dr^2$ are exactly
the same as in the vacuum case.)

We shall now apply Stokes' theorem for a suitably chosen
$2$-dimensional vector in order to prove that
there exist no regular, non-trivial solutions to eq. (\ref{MAT2})
with asymptotic behavior (\ref{r2h}).
Consider the vector
\be
w \, = \, \rho \, \nabdel \, e^{- \hphi}
\label{w}
\ee
in the $(\rho, z)$ plane
($\nabdel = (\partial_{\rho}, \partial_z)$).
Stokes' theorem for $w$ and the domain
${\cal S}$ with counter clockwise oriented
boundary $\partial {\cal S}$ yields
\be
\oint_{\partial {\cal S}} \rho \, e^{- \hphi} \,
(\hphi,_z d \rho - \hphi,_{\rho} dz) \, = \,
\int_{{\cal S}} \rho \, e^{- \hphi} \, [ \,
|\nabdel \hphi|^2 \, - \,
(\frac{1}{\rho} \hphi,_{\rho} + \Delta^{(\delta)} \hphi) \, ] \;
d \rho \, dz \, .
\label{Green}
\ee
The crucial observation consists in the fact that
the r.h.s. of this identity is a sum of two
non-negative terms, provided that the metric
$G_{AB}$ of the target manifold $N$ is Riemannian.
This is immediately seen from eqs. (\ref{MAT2})
and (\ref{lapphi}), which together yield
\be
- \, (\hphi,_{\rho} + \rho \Delta^{(\delta)} \hphi)
\, = \,
\rho \, G_{AB}(\phi) \, \phi^A,_z \phi^B,_z \, \geq \, 0 \, .
\label{pos}
\ee
Our last task is to show that the boundary integral on the
l.h.s. of eq. (\ref{Green}) vanishes, which then implies
that $\hphi$ is constant.
Using prolate spheroidal co-ordinates, we
have to demonstrate that $\lim_{R \rightarrow \infty} I_R = 0$,
where
\be
I_R \, = \, \mu \, \oint_{\partial {\cal S}_R}
e^{- \hphi} \,
[ \, (1 - y^2) \, \hphi,_y dx \, - \, (x^2 - 1) \, h,_x dy \, ] \, .
\label{Ic}
\ee
Here the oriented boundary $\partial {\cal S}_R$ of the
domain of outer communications is the rectangle
$\gamma^1_R = \{ y = 1, x = R...1 \}$,
$\gamma^2 = \{ x = 1, y = 1...-1 \}$,
$\gamma^3_R = \{ y = -1, x = 1...R \}$ and
$\gamma^4_R = \{ x = R, y = -1...1 \}$.
The finiteness of the Ricci scalar and the regularity
of the derivatives of $\hphi$ with respect to
Boyer-Lindquist co-ordinates imply that
$\hphi,_x$, $\hphi,_y$ and $exp(- \hphi)$ remain finite
along $\gamma^1_R$, $\gamma^2$ and $\gamma^3_R$.
Hence, both integrals in eq. (\ref{Ic})
vanish along these parts of the boundary.
It remains to consider the contribution
from the integration
along $\gamma^4_R$ as $R \rightarrow \infty$, that is
\be
\lim_{R \rightarrow \infty} I_R \, = \,
- \mu \, \lim_{R \rightarrow \infty} \int_{\gamma^4_R}
e^{- \hphi} \,  (x^2 - 1) \, h,_x dy \, = \,
- \int_0^{\pi} \lim_{r \rightarrow \infty} \,
(e^{- \hphi} r^2 h,_r) \, \sin \vartheta \, d \vartheta \, .
\label{inti}
\ee
This integral vanishes as well,
as a consequence of the asymptotic behavior (\ref{r2h}) of $r^2 h,_r$.
Thus, the l.h.s. of eq. (\ref{Green})
is zero, implying that both non-negative integrands on the r.h.s.,
$|\nabdel \hphi|^2$ and $-[\hphi,_{\rho} + \rho \Delta \hphi]$,
vanish. Hence, $\hphi$ is constant in the entire domain
of outer communications and, since
$\lim_{r \rightarrow \infty} \hphi = 0$,
we finally have
\be
\hphi \, = \, 0 \, ,
\label{h0}
\ee
and the metric
(\ref{Kerrds}) reduces to the ordinary Kerr metric.
Eventually $\hphi = 0$ implies that
$\phi$ is a constant map.

To summarize, we have shown that the uniqueness result
previously obtained in \cite{HSU} for the non-rotating case
generalizes to rotating black holes: \\

{\em{Let $(M,g)$ be an asymptotically flat,
stationary and axisymmetric spacetime
with Killing fields $k$ and $m$.
Let $\phi$ denote a mapping with harmonic action from $(M,g)$
into a Riemannian manifold $(N,G)$, $\phi$ being invariant
under the action of $k$ and $m$.
Then the only stationary and axisymmetric,
asymptotically flat black hole solution
of the coupled Einstein-matter equations with regular
event horizon consists of the Kerr metric and a constant map
$\phi_0$}}. \\

To conclude, we note that this result, combined
with the corresponding theorem for non-rotating
black holes mentioned above and the strong rigidity theorem,
implies that the Kerr metric is the unique stationary,
asymptotically flat black hole solution of self-gravitating
harmonic mappings with Riemannian target manifolds.
It is also worth pointing out that the generalization
of this result to the case where electromagnetic fields are
taken into account as well is straightforward.
In this case, the analogous no-hair theorem applies to the
Kerr-Newman metric, being the unique electrovac black hole
solution with selfgarvitating harmonic mappings.


%
\end{document}